# Neural Gas based classification of Globular Clusters

Giuseppe Angora[1][0000−0002−0316−6562], Massimo Brescia[2][0000−0001−9506−5680], Stefano Cavuoti[1,2,3][0000−0002−3787−4196], Giuseppe Riccio[2][0000−0001−7020−1172], Maurizio Paolillo[1][0000−0003−4210−7693], and Thomas H. Puzia[4][0000−0003−0350−7061]

[1] University of Naples Federico II - Dept. of Physics "E. Pancini", via Cintia 6, I-80135 Napoli, Italia
`gius.angora@gmail.com`
[2] INAF - Astronomical Observatory of Capodimonte, via Moiariello 16, I-80131 Napoli, Italia
[3] INFN - Napoli Unit, via Cintia 6, I-80135 Napoli, Italia
[4] Institute of Astrophysics, Pontificia Universidad Catolica de Chile, Av. Vicua Mackenna 4860, Macul, Santiago, Chile

**Abstract.** Within scientific and real life problems, classification is a typical case of extremely complex tasks in data-driven scenarios, especially if approached with traditional techniques. Machine Learning supervised and unsupervised paradigms, providing self-adaptive and semi-automatic methods, are able to navigate into large volumes of data characterized by a multi-dimensional parameter space, thus representing an ideal method to disentangle classes of objects in a reliable and efficient way.

In Astrophysics, the identification of candidate Globular Clusters through deep, wide-field, single band images, is one of such cases where self-adaptive methods demonstrated a high performance and reliability. Here we experimented some variants of the known Neural Gas model, exploring both supervised and unsupervised paradigms of Machine Learning for the classification of Globular Clusters. Main scope of this work was to verify the possibility to improve the computational efficiency of the methods to solve complex data-driven problems, by exploiting the parallel programming with GPU framework. By using the astrophysical playground, the goal was to scientifically validate such kind of models for further applications extended to other contexts.

**Keywords:** data analytics, astroinformatics, globular clusters, machine learning, neural gas

## 1 Introduction

The current and incoming astronomical synoptic surveys require efficient and automatic data analytics solutions to cope with the explosion of scientific data amounts to be processed and analyzed. This scenario, quite similar to other scientific and social contexts, pushed all communities involved in data-driven disciplines to explore data mining techniques and methodologies, most of which connected to the Machine Learning (hereafter ML) paradigms, i.e. supervised and unsupervised self-adaptive learning and parameter space optimization [1, 2]. Following this premise, this paper is focused on the investigation about the use of a particular kind of ML methods, known as Neural Gas (NG) models [3], to solve classification problems within the astrophysical context, characterized by a complex multi-dimensional parameter space. In order to scientifically validate such models, we decided to approach a typical astrophysical playground, already solved with ML methods [4–6] and to use in parallel other two ML techniques, chosen among the most standard, respectively, Random Forest [7] and Multi Layer Perceptron Neural Network [8], as comparison baseline. The astrophysical case is related to the identification of Globular Clusters (GCs) in the galaxy NGC1399 using single band photometric data obtained through observations with the Hubble Space Telescope (HST) [4, 9, 10]. The physical identification and characterization of a Globular Cluster (GC) in external galaxies is considered important for a variety of astrophysical problems, from the dynamical evolution of binary systems, to the analysis of star clusters, galaxies and cosmological phenomena [10]. Here, the capability of ML methods to learn and recognize peculiar classes of objects, in a complex and noising parameter space and by learning the hidden correlation among objects parameters, has been demonstrated particularly suitable in the problem of GC classification [4]. In fact, multi-band wide-field photometric data (colours and luminosities) are usually required to recognize GCs within external galaxies, due to the high risk of contamination of background galaxies, which appear indistinguishable from galaxies located few Mpc away, when observed by ground-based instruments. Furthermore, in order to minimize the contamination, high-resolution space-borne data are also required, since they are able to provide particular physical and structural features (such as concentration, core radius, etc.), thus improving the GC classification performance [9]. In [4] we demonstrated the capability of ML methods to classify GCs using only single band images from Hubble Space Telescope with a classification accuracy of 98.3%, a completeness of 97.8% and only 1.6% of residual contamination. Thus confirming that ML methods may yield low contamination by minimizing the observing requirements and extending the investigation to the outskirts of nearby galaxies. These results gave us an optimal playground where to train NG models and to validate their potential to solve classification problems characterized by complex data with a noising parameter space. The paper is structured as follows: in Sec. 2 we describe the data used to test various methods. In Sec. 3 we provide a short methodological and technical description of the models. In Sec. 4 we describe the experiments and results about the parameter space analysis and classification experiments, while in Sec. 5 we discuss the results and draw our conclusions. This paper is an extended version of the work presented at the DAMDID 2017 Conference and published in the related Proceedings [11]. It includes three new sections, respectively, 3.6, 4.3 and 5.2, plus two new figures (Fig. 1 and 4) and the re-phrased abstract section, whose total amount is more than 30% of the previous work [11].

## 2  The Astrophysical Playground

The HST single band data are very suitable to investigate the classification of GCs. They, in fact, represent deep and complete in terms of wide field coverage i.e. able to sample the GC population, to ensure a high S/N ratio required to measure structural parameters [12]. Furthermore, they provide the possibility to study the overall properties of the GC populations, which usually may differ from those of the central region of a galaxy. With such data we intend to verify that Neural Gas based models could be able to identify GCs with low contamination even with single band photometric information. Throughout the confirmation of such behavior, we are confident that these models could solve other astrophysical problems as well as other data-driven problem contexts.

## 2.1 The data

The data used in the described experiment consist of wide field single band HST observations of the giant elliptical NGC1399 galaxy, located in the core of the Fornax cluster [10]. Due to its distance (D=20.130 Mpc, see [13]), it is considered an optimal case where to cover a large fraction of its GC system with a restricted number of observations. This dataset was used by [9] to study the GC-LMXB connection and the structural properties of the GC population. The optical data were taken with the HST Advanced Camera for Surveys, in the broad V band filter, with 2108 seconds of integration time for each field. The detection of GCs relies basically upon two aspects: the shape of the image (which differs from the instrumental PSF) and the colours (i.e. the ratio of observed fluxes at different wavelengths). The shape allows to disentangle large systems from stars (which are PSF-like), while the colours are needed to disentangle GCs from other extended systems, such as background galaxies (Fig. 1).

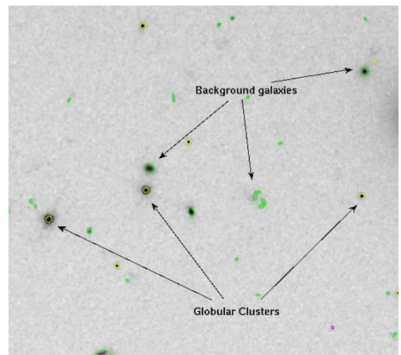

**Fig. 1.** A sub-section of the HST (Hubble Space Telescope) image used to build the dataset for the experiment. It is obtained with the ACS (Advanced Camera for Survey) used to detect Globular Clusters (GCs) around N1399. GCs (in yellow) are difficult to distinguish from background galaxies (in green), based only on single band images.

The source catalog was generated using Sextractor [14], [15], by imposing a minimum area of 20 pixels: it contains 12915 sources and reaches $7\sigma$ detection at m V=27.5, i.e. 4 mag below the GC luminosity function, thus allowing to sample the entire GC population (see [4] for details). The source subsample used to build our Knowledge Base (KB) to train the ML models, is composed by 2100 sources with 11 features (7 photometric and 4 morphological parameters). Such parameter space includes three aperture magnitudes within 2, 6 and 20 pixels (*mag aper1, mag aper2, mag aper3* ), isophotal magnitude (*mag iso*), kron radius (*kron rad*) central surface brightness (*mu0* ), FWHM (*fwhm im*), and the four structural parameters, respectively, *ellipticity,* King's tidal, effective and core radii (*calr t, calr h, calr c*).

The target values of the KB required as ground truth for training and validation, i.e. the binary column indicating the source as GC or not GC, is provided through the typical selection based on multi-band magnitude and colour cuts. The original 2100 sources having a target assigned have been randomly shuffled and split into a training (70%) and a blind test set (30%).

## 3    The Machine Learning Models

In our work we tested three different variants of the Neural Gas model, using two additional machine learning methods, respectively feed-forward neural network and Random Forest, as comparison benchmarks. In the following all main features of these models are described.

### 3.1    Growing Neural Gas

Growing Neural Gas (GNG) is presented by [16] as a variant of the Neural Gas algorithm (introduced by [3]), which combines the Competitive Hebbian Learning (CHL, [17]) with a vector quantization technique to achieve a learning that retains the topology of the dataset. Vector quantization techniques [17] encode a data manifold, e.g. $V \subseteq R^m$, using a finite set of reference vectors $w = w_1 \ldots w_N, w_i \in R^m, i = 1 \ldots N$. Every data vector $v \in V$ is described by the best matching reference vector $w_{i(v)}$ for which the distortion error $d(v, w_{i(v)})$ is minimal. This procedure divides the manifold $V$ into a number of subregions: $V_i = \{v \in V : ||v - w_i|| \leq ||v - w_j|| \; \forall j\}$, called Voronoi polyhedra [18], within which each data vector v is described by the corresponding reference vector $w_i$. The Neural Gas network is a vector quantization model characterized by N neural units, each one associated to a reference vector, connected to each other. When an input is extracted, it induces a synaptic excitation detected by all the neurons in the graph and causes its adaptation. As shown in [3], the adaptation rule can be described as a "winner-takes-most" instead of "winner-takes-all" rule:

$$\Delta w_i = \varepsilon h_\lambda(v, w_i) \cdot (v - w_i), \qquad i = 1 \ldots N. \tag{1}$$

The step size describes the overall extent of the adaptation. While $h_\lambda(v, w_i) = h_\lambda(k_i(v, w))$ is a function in which $k_i$ is the "neighborhood-ranking" of the reference vectors. Simultaneously, the first and second Best Matching Units (BMUs) develop connections between each other [3]. Each connection has an "age"; when the age of a connection exceeds a pre-specified lifetime T, it is removed [3]. Martinez's reasoning is interesting [17]: they demonstrate how the dynamics of neural units can be compared to a gaseous system. Let's define the density of vector reference at location u through $\rho(u) = F^{-1}_{BMU(u)}$, where $F_{BMU(u)}$ is the volume of Voronoi Polyedra. Hence, $\rho(u)$ is a step function on each Voronoi polyhedra, but we can still imagine that their volumes change slowly from one polyhedra to the next, with $\rho(u)$ continuous. In this way, it is possible to derive an expression for the average change:

$$\Delta w_i \propto \frac{1}{\rho^{1+2/m}}\left[\partial_u P(u) - \frac{2+m}{m}\frac{P}{\rho}\partial_u \rho(u)\right] \tag{2}$$

where $P(u)$ is the data point distribution. The equation suggests the name Neural Gas: the average change of the reference vectors corresponds to a motion of particles in a potential $V(u) = -P(u)$. Superimposed on the gradient of this potential there is a force proportional to $-\partial_u \rho(u)$, which points toward the direction of the space where the particle density is low. Main idea behind the GNG network is to successively add new units to an initially small net- work, by evaluating local statistical measures collected during previous adaptation steps [16].

Therefore, each neural unit in the graph has associated a local reconstruction error, updated for the BMU at each iteration (i.e. each time an input is extracted): $\Delta error_{BMU} = //w_{BMU} - v//$. Unlike the Neural Gas network, in the GNG the synaptic excitation is limited to the receptive fields related to the Best Matching Unit and its topological neighbors: $\Delta w_i = E_i(v - w_i)$, $i \in (BMU, n)$, $\forall n \in neighbours(BMU)$. It is no longer necessary to calculate the ranking for all neural units, but it is sufficient to determine the first and the second BMU. The increment of the number of units is performed periodically: during the adaptation steps the error accumulation allows to identify the regions in the input space where the signal mapping causes major errors. Therefore, to reduce this error, new units are inserted in such regions [16]. An elimination mechanism is also provided: once the connections, whose age is greater than a certain threshold, have been removed, if their connected units remain isolated (i.e. without emanating edges), those units are removed [16].

### 3.2 GNG with Radial Basis Function

Fritzke describes an incremental Radial Basis Function (RBF) network suitable for classification and regression problems [16]. The network can be figured out as a standard RBF network [19], with a GNG algorithm as embedded clustering method, used to handle the hidden layer. Each unit of this hybrid model (here- after GNGRBF) is a single perceptron with an associated reference vector and a standard deviation. For a given input-output pair $(v, y)$, $v \in R^n$, $y \in R^m$, the activation of the $i$-th unit is described by $D_i(v) = e^{-//v - w_i// / \sigma_i}$. Each of the single perceptron computes a weighted sum of the activations: $O_i = \Sigma_j w_{ij} D_j(v)$, $i = 1 \dots m$. The adaptation rule applies to both reference vectors forming the hidden layer and the RBF weights. For the first, the adaptation rule is the same of the updating rule for the GNG network, while for the weights:

$$\Delta w_{ij} = \eta D_j(y_i - o_i), i = 1 \dots m, j \in N. \tag{3}$$

Similarly to the GNG network, new units are inserted where the prediction error is high, updating only the Best Matching Unit at each iteration: $\Delta error_{BMU} = \Sigma_i y_i - O_i$.

### 3.3 Supervised Growing Neural Gas

The Supervised Growing Neural Gas (SGNG) algorithm is a modification of the GNG algorithm that uses class labels of data to guide the partitioning of data into optimal clusters [20], [21]. Each of the initial neurons is labelled with a unique class label. To reduce the class impurity inside the cluster, the original learning rule 1 is reformulated by considering the case where the BMU belongs or not to the same class of the neuron whose reference vector is the closest to the current input. Depending on such situation the SGNG learning rule is expressed alternatively as:

$$\begin{cases} \Delta w_n = -\varepsilon \frac{v - w_n}{\|v - w_n\|} & or \\ \Delta w_n = +\varepsilon \frac{v - w_n}{\|v - w_n\|} + repulsion(sn, n) \end{cases} \tag{4}$$

where *sn* is the nearest class neuron and *repulsion*() is a function specifically introduced to maintain neurons sufficiently distant one each other. For the neuron that is topologically close to the neuron *sn*, the rule intends to increase the clustering accuracy [21]. The insertion mechanism has to reduce not only the intra-distances between data in a cluster, but also the impurity of the cluster. Each unit has associated two kinds of error: an aggregated and a class error. A new neuron is inserted close to the neuron having a highest class error accumulated, while the label is the same as the neuron label with the greater aggregated error.

### 3.4  Multi Layer Perceptron

The Multi Layer Perceptron (MLP) architecture is one of the most typical feed-forward neural networks [8]. The term feed-forward is used to identify basic behavior of such neural models, in which the impulse is propagated always in the same direction, e.g. from neuron input layer towards output layer, through one or more hidden layers (the network brain), by combining the sum of weights associated to all neurons. As easy to understand, the neurons are organized in layers, with proper own role. The input signal, simply propagated throughout the neurons of the input layer, is used to stimulate next hidden and output neuron layers. The output of each neuron is obtained by means of an activation function, applied to the weighted sum of its inputs. The weights adaptation is obtained by the Logistic Regression rule [22], by estimating the gradient of the cost function, the latter being equal to the logarithm of the likelihood function between the target and the prediction of the model. In this work, our implementation of the MLP is based on the public library Theano [23].

### 3.5  Random Forest

Random Forest (RF) is one of the most widely known machine learning ensemble methods [7], since it uses a random subset of candidate data features to build an ensemble of decision trees. Our implementation makes use of the public library scikit-learn [24]. This method has been chosen mainly because it provides for each input feature a score of importance (rank) measured in terms of its informative contribution percentage to the classification results. From the architectural point of view, a RF is a collection (forest) of tree-structured classifiers $h(x, \theta_k)$, where the $\theta_k$ are independent, identically distributed random vectors and each tree casts a unit vote for the most popular class at input. Moreover, a fundamental property of the RF is the intrinsic absence of training overfitting [7].

### 3.6  Implementation details

In this work we compare the classification results and computational performance referred to three different implementations: besides the original networks (GNG$_{old}$, GNGRBF$_{old}$), we developed different versions, including one exclusively using the numpy methods [25] in the case of GNG, and another completely written using theano library [23] in all cases of GNG, GNGRBF and MLP. The MLP model was selected as test bench to compare NG-based models with the most general-purpose feedforward neural network. Concerning the *old* version of NG methods, main modification in the new versions is the random extraction of a sample batch, whose size varies between 1 (corresponding to the *old* case) and the total number of samples. The advantages are: (*i*) the method allows to use the matrices algebra; and (*ii*) during the learning phase the adaptation of the weights and the reference vectors reflect the complexity of the data set. The use of the theano library allows the possibility to compute the gradient of a cost function depending on the parameters available. For the GNG the cost function could be represented by the quantization error, which is a measure of similarity among the samples allocated by the same BMU. Given a data set $v \in V$ composed by $/V/$ records, distributed among $p$ neurons, the quantization error is given by:

$$QE = \frac{1}{2|V|} \sum_{i=1}^{p} \sum_{n \in BMU_i} ||v_n - w_n||^2$$

In addition, the reference vectors are adapted with:

$$\Delta \mathbf{W} = -\eta \, \nabla_{\mathbf{W}}(\mathrm{QE}),$$

where $\mathbf{W}$ is the matrix of the reference vectors, $\eta$ is the learning rate and $\nabla_{\mathbf{W}}$ is the gradient derived from reference vectors. For the supervised networks (GNGRBF$_{theano}$ and MLP) the cost function is computed as the negative logarithm of the likelihood function between the target and the output:

$$C = \frac{1}{|V|} \mathcal{L}(\theta = \{\mathbf{W}, b\}, V) = \frac{1}{|V|} \sum_{i=0}^{|V|} \log(P(y = y_i | \mathbf{X}_i, \mathbf{W}, b)) \tag{5}$$

## 4   The experiments

The five models previously introduced have been applied to the dataset described in Sec. 2.1 and their performances have been compared to verify the capability of NG models to solve particularly complex classification problems, like the astrophysical identification of GCs from single-band observed data.

### 4.1   The Classification Statistical Estimators

In order to evaluate the performances of the selected classifiers, we decided to use three among the classical and widely used statistical estimators, respectively, average efficiency, purity, completeness and F1-score, which can be directly derived from the confusion matrix [26]. The average efficiency (also known as accuracy, hereafter AE), is the ratio between the sum of correctly classified objects on both classes (true positives for both classes, hereafter tp) and the total amount of objects in the test set. The purity (also known as precision, hereafter pur) of a class measures the ratio between the correctly classified objects and the sum of all objects assigned to that class (i.e. tp⁄[tp+fp], where fp indicates the false positives). While the completeness (also known as recall, hereafter comp) of a class is the ratio tp⁄[tp+fn], where fn is the number of false negatives of that class. The quantity tp+fn corresponds to the total amount of objects belonging to that class. The F1-score is a statistical test that considers both the purity and completeness of the test to compute the score (i.e. 2[pur·comp]⁄[pur+comp]). By definition, the dual quantity of the purity is the contamination, another important measure which indicates the amount of misclassified objects for each class. In statistical terms, it is well known the classical tradeoff between purity and completeness in any classification problem, particularly accentuated in astrophysical problems [27]. In the specific case of the GC identification, from the astrophysical point of view, we were mostly interested to the purity, i.e. to ensure the highest level of true GCs correctly identified by the classifiers [4]. However, within the comparison experiments described in this work, our main goal was to evaluate the performances of the classifiers mostly related to the best tradeoff between purity and completeness.

## 4.2   Analysis of the Data Parameter Space

Before to perform the classification experiments, we preliminarily investigated the parameter space, defined by the 11 features defined in Sec. 2.1, identifying each object within the KB dataset of 2100 objects. Main goal of this phase was to measure the importance of any feature, i.e. its relevance in terms of informative contribution to the solution of the problem. In the ML context, this analysis is usually called feature selection [28]. Its main role is to identify the most relevant features of the parameter space, trying to minimize the impact of the well known problem of the curse of dimensionality, i.e. the fact that ML models exhibit a decrease of performance accuracy when the number of features is significantly higher than optimal [29]. This problem is mainly addressed to cases with a huge amount of data and dimensions. However, its effects may also impact contexts with a limited amount of data and parameter space dimension. The Random Forest model resulted particularly suitable for such analysis, since it is intrinsically able to provide a feature importance ranking during the training phase. The feature importance of the parameter space, representing the dataset used in this work, is shown in Fig. 2. From the astrophysical point of view, this ranking is in accordance with the physics of the problem. In fact, as expected, among the five most important features there are the four magnitudes, i.e. the photometric log-scale measures of the observed object's photonic flux through different apertures of the detector. Furthermore, almost all photometric features resulted as the most relevant. Finally, by looking at the Fig. 2, there is an interesting gap between the first six and the last five features, whose cumulative contribution is just ∼ 11% of the total. Finally, a very weak joined contribution (∼3%) is carried by the two worst features (*kron_rad* and *calr_c*), which can be considered as the most noising or redundant features for the problem domain.

**Table 1.** Statistical analysis of the classification performances obtained by the five ML models on the blind test set for the four selected experiments. All quantities are expressed in percentage and related to average efficiency (AE), purity for each class (purGC, purNotGC), completeness for each class (compGC, compNotGC) and the F1-score for GC class. The contamination is the dual value of the purity. The numerical index of used features corresponds to the list in Fig. 2.

| ID | features | Estimator | RF% | MLP% | SGNG% | GNGRBF% | GNG% |
|----|----------|-----------|------|------|-------|---------|------|
| E1 | 1,2,3,5 | AE | 88.9 | 84.4 | 88.1 | 88.1 | 88.4 |
|    |          | purGC | 85.9 | 80.1 | 89.7 | 85.4 | 83.7 |
|    |          | compGC | 87.3 | 82.6 | 80.3 | 85.7 | 89.2 |
|    |          | F1-scoreGC | 86.6 | 81.3 | 84.7 | 85.5 | 86.4 |
|    |          | purNotGC | 91.0 | 87.6 | 87.2 | 90.0 | 92.1 |
|    |          | compNotGC | 89.7 | 85.6 | 93.0 | 89.6 | 88.1 |
| E2 | 1,2,3,4,5,6 | AE | 89.0 | 85.1 | 87.3 | 88.3 | 83.2 |
|    |          | purGC | 84.9 | 77.0 | 81.0 | 82.9 | 74.0 |
|    |          | compGC | 89.2 | 90.7 | 90.3 | 90.0 | 91.1 |
|    |          | F1-scoreGC | 87.0 | 83.3 | 85.4 | 86.3 | 81.7 |
|    |          | purNotGC | 92.2 | 92.6 | 92.7 | 92.6 | 92.6 |
|    |          | compNotGC | 89.0 | 85.6 | 85.7 | 87.4 | 80.0 |
| E3 | 1,2,3,4,5,6,10 | AE | 89.0 | 83.2 | 85.1 | 89.2 | 86.8 |
|    |          | purGC | 85.2 | 77.2 | 80.0 | 86.0 | 84.1 |
|    |          | compGC | 88.8 | 83.8 | 84.9 | 88.0 | 83.8 |
|    |          | F1-scoreGC | 87.0 | 80.4 | 82.4 | 87.0 | 83.9 |
|    |          | purNotGC | 91.9 | 88.0 | 89.0 | 91.5 | 88.7 |
|    |          | compNotGC | 89.9 | 83.2 | 85.1 | 89.8 | 88.4 |
| E4 | 1,2,3,4,5,6,7,8,9 | AE | 89.5 | 86.0 | 88.1 | 88.7 | 83.8 |
|    |          | purGC | 85.3 | 82.5 | 84.1 | 83.8 | 78.3 |
|    |          | compGC | 90.0 | 83.8 | 87.6 | 90.0 | 83.8 |
|    |          | F1-scoreGC | 87.6 | 83.1 | 85.8 | 86.8 | 81.0 |
|    |          | purNotGC | 92.7 | 88.6 | 91.1 | 92.6 | 88.1 |
|    |          | compNotGC | 89.1 | 87.5 | 88.1 | 88.2 | 84.1 |

Based on such considerations, the analysis of the parameter space provides a list of most interesting classification experiments to be performed with the selected five ML models. This list is reported in Table 1. The experiment E1 is useful to verify the efficiency by considering the four magnitudes. The experiment E2 is based on the direct evaluation of the best group of features as derived from the importance results. The classification efficiency of the full photometric subset of features is evaluated through the experiment E3. Finally, the experiment E4 is performed to verify the results by removing only the two worst features.

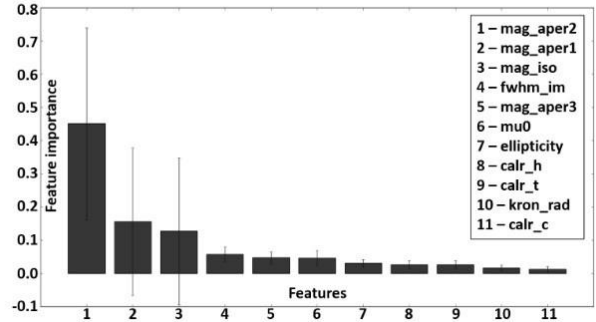

**Fig. 2.** The feature importance ranking obtained by the Random Forest on the 11-feature domain of the input dataset during training (see Sec. 2.1 for details). The vertical lines report the importance estimation error bars.

### 4.3 The classification of Globular Clusters

By applying the feature space analysis, the original parameter domain has been simplified, reducing the number of features used. The classification experiments have been performed on the data, presented in Sec. 2.1, composed by 2100 objects having up to a maximum of 9 features as parameter space (Table 1). The dataset has been randomly shuffled and split into a training set of 1470 objects (70% of the whole KB) and a blind test set of 630 objects (the residual 30% of the KB). These datasets have been used to train and test the selected five ML classifiers. The analysis of results, reported in Table 1, has been performed on the blind test set, in terms of the statistical estimators defined in Sec. 4.2. Besides the classification quality evaluation, the three models and their variants have been also investigated in terms of computing efficiency. In fact, NG-based models are not intrinsically scalable with the data volume, but they require a careful optimization. The *old* versions of GNG and GNGRBF have the strong limitation of the absence of any batch system, being forced to evaluate one by one the data entries during the learning phase. Therefore, a direct comparison of computing time vs data size among *old* and evolved batch-based variants is intrinsically in favor of the latter at least by one order of magnitude. However, an equally interesting comparison involving *old* and evolved versions of the models is shown in the two top panels of Fig. 4, where are evaluated, respectively, the three variants of the GNG model in terms of execution time as function of the quantization error along the training process and the two versions of GNGRBF vs MLP in terms of execution time as function of the Root Mean Square (RMS) of the learning error along the training process. While a comparison of the computing efficiency, in terms of execution time as function of training data increasing size is reported in the two bottom panels of Fig. 4, showing, respectively, a direct evaluation among all NG-based models and a direct comparison among the *theano*-based version for all three types of ML models. The incremental dataset was obtained as multiple repetitions of the original dataset with the addition of white noise.

## 5   Discussion

As already underlined, main goal of this work was the validation of NG models as efficient classifiers in noising and multi-dimensional problems, with performances at least comparable to other ML methods, considered traditional in terms of their use in such kind of problems.

### 5.1   Analysis of classification performance

By looking at Table 1 and focusing on the statistics for the three NG models, it is evident that their result is able to identify GCs from other background objects, reaching a satisfying tradeoff between purity and completeness in all experiments and for both classes. The occurrence of statistical fluctuations is mostly due to the different parameter space used in the four experiments. Nevertheless, none of the three NG models overcome the others in terms of the measured statistics. If we compare the NG models with the two additional ML methods (Random Forest and MLP neural network), their performances appears almost the same. This implies that NG methods show classification capabilities fully comparable to other ML methods.

**Table 2.** Statistics for the three NG models related to the common predictions of the correctly classified objects. Second column is referred to both classes, while the third and fourth columns report, respectively, the statistics for single classes.

| EXP ID | GC+notGC% | GC% | notGC% |
|--------|-----------|------|--------|
| E1 | 86.0 | 85.4 | 86.9 |
| E2 | 79.8 | 79.8 | 79.8 |
| E3 | 81.1 | 82.5 | 79.2 |
| E4 | 77.8 | 77.4 | 78.4 |

Another interesting aspect is the analysis of the degree of coherence among the NG models in terms of commonalities within classified objects. Table 2 reports the percentages of common predictions for the objects correctly classified by considering, respectively both and single classes. On average, the three NG models are in agreement among them for about 80% of the objects correctly classified.

This is also confirmed by looking at the Fig. 3, where the tabular results of Table 2 are showed through the Venn diagrams, reporting also more details about their classification commonalities. Finally, from the computational efficiency point of view, the NG models have theoretically a higher complexity than Random Forest and neural networks. But, since they are based on a dynamic evolution of the internal structure, their complexity strongly depends on the nature of the problem and its parameter space. Nevertheless, all the presented ML models have a variable architectural attitude to be compliant with the parallel computing paradigms. Besides the embarrassingly parallel architecture of the Random Forest, the use of optimized libraries, like Theano [23], make also models like MLP highly efficient. From this point of view NG models have a high potentiality to be parallelized. By optimizing GNG, the GNGRBF would automatically benefit, since both share the same search space, except for the RBF training additional cost.

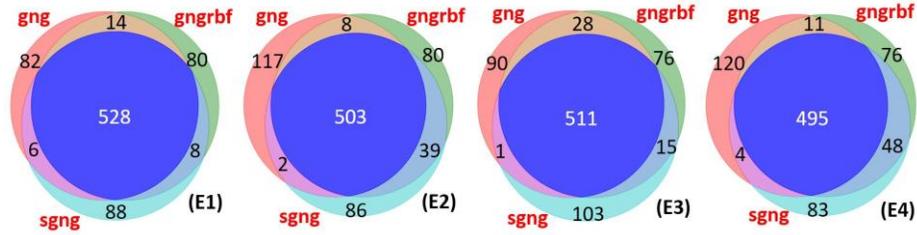

**Fig. 3.** The Venn diagram related to the prediction of all (both GCs and not GCs) correctly classified objects performed by the three Neural Gas based models (GNG, GNGRBF and SGNG) for the experiments, respectively, E1 (a), E2 (b), E3 (c) and E4 (d). The intersection areas (dark grey in the middle) show the objects classified in the same way by different models. Internal numbers indicate the amount of objects correctly classified for each sub-region.

In practice, the hidden layer of the supervised network behaves just like a GNG network whose neurons act as inputs for the RBF net- work. Consequently, with the same number of iterations, the GNGRBF network performs a major number of operations. On the other hand, the SGNG network is similar to the GNG network, although characterized by a neural insertion mechanism over a long period, thus avoiding too rapid changes in the number of neurons and excessive oscillations of reference vectors. Therefore, on average, the SGNG network computational costs are higher than the models based on the standard Neural Gas mechanism.

### 5.2 Analysis of computing performance

From the computing efficiency point of view, all NG-based models have a complexity of the order $O(D \times N \times M)$, with D dimensions of the parameter space, N amount of data and M number of neurons. But, since they are based on a dynamic evolution of the internal structure, their complexity strongly depends on the nature of the problem and its parameter space. In Fig. 4 (top left panel) we can see how the different architectures influence the computing time of the same model. In fact, different ways to store the memory induce different *loading times*. The *old* implementations having a simpler data structure are, in fact, faster at the beginning of the training than the *theano* implementations. During the training such initial gap is filled by means of the theano optimization, while at the end of the process we obtain a final quantization error which is similar but reached at different times: the *old* implementation is the slower while *numpy* and *theano* are comparable. This should be even more remarkable with a wider dataset and we plan to perform such kind of analysis in the next future. By looking at the top right panel of Fig. 4 we can see the trend of time as function of RMS for the supervised models (i.e. GNGRBF$_{theano}$, GNGRBF$_{old}$ and MLP$_{theano}$). Between the two versions of GNGRBF it appears an initial gap. After that, the GNGRBF implementation needs a sufficient number of neurons to codify the input data in order to reduce the RMS error. This is clearly connected to the use of the *batch system* in the implementation. Once out from the local minimum the *theano* implementation is close to the computing performance of the *old* implementation, reaching the final solution in less time.

As in the previous case we expect that the improvement should be more evident with a larger dataset. We noticed that MLP has a completely different behavior, by immediately reaching a good solution, spending much of the computing time to slightly improve the learning quality. From what computing time vs data batch size concerns, the two bottom panels of Fig. 4 report the execution time required by different models as function of the growing batch size of input data. As evident in the bottom left panel of Fig. 4, the *theano* version of GNG is outperforming its *numpy* version.

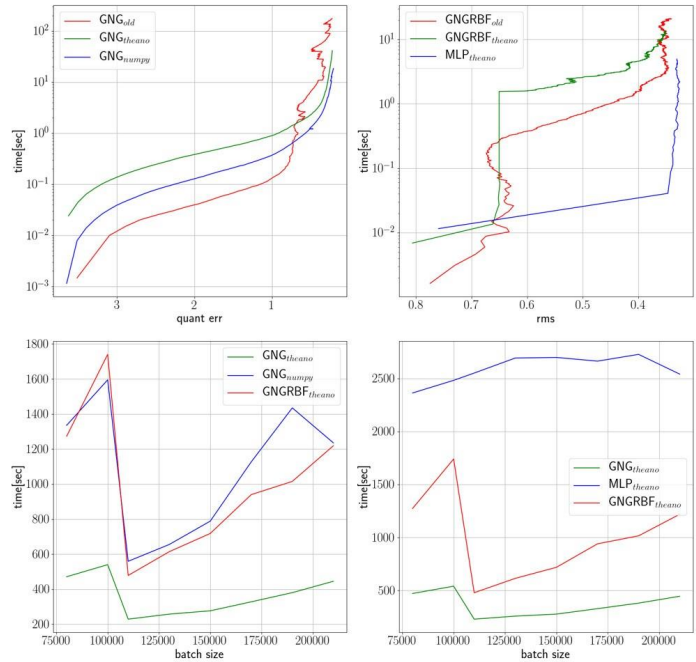

**Fig. 4.** Top Left panel: execution time (log scale) as a function of the quantization training error for the three GNG networks: GNG$_{old}$ (red line), GNG$_{theano}$ (green line), GNG$_{numpy}$ (blue line). Top Right panel: Execution time (log scale) as a function of the root mean square of the training error for the three supervised networks: GNGRBF$_{old}$ (red line), GNGRBF$_{theano}$ (green line), MLP (blue line). Bottom panels: execution time vs increasing data size for, respectively, the three NG-based models (left) and the three *theano*-based models (right). GNGRBF$_{theano}$ (red line), GNG$_{theano}$ (green line), while blue line is used on the left for GNG$_{numpy}$ and on the right for the MLP.

The almost same trend between GNG$_{numpy}$ and GNGRBF$_{theano}$ can be easily motivated by taking into account the higher computational cost of RBF network as well as an increased number of calculations imposed by the hybrid model, which shades the improvements carried by the *theano* optimization. The bottom right panel of Fig. 4 poses a direct comparison among the three *theano* versions of the three ML models. As in the previous case, the GNG model outperforms the others. It is also remarkable the behavior of MLP, which shows an almost uniform trend, motivated by considering its intrinsic learning mode, less sensible to the variation of data batch size. While a usual sharp decay is again present in the behavior of GNGRBF model.

## 6 Conclusions

In conclusion, although a more intensive test campaign on these models is still ongoing, we can assert that Neural Gas based models are very promising as problem-solving methods, also in presence of complex and multi-dimensional classification and clustering problems, especially if preceded by an accurate analysis and optimization of the parameter space within the problem domain.

### Acknowledgements

MB acknowledges the *INAF PRIN-SKA 2017 program 1.05.01.88.04* and the funding from *MIUR Premiale 2016: MITIC*.